\documentclass[a4paper,12pt]{article}
\usepackage[english]{babel}

\def\beq{\begin{equation}}
\def\eeq{\end{equation}}
\def\bea{\begin{eqnarray}}
\def\eea{\end{eqnarray}}
\def\zbar{\bar{z}}

\def\calc{\hat {\cal C}}
\def\cald{\hat {\cal D}}
\def\vac{\left| 0 \right>}

\begin{document}

\pagestyle{plain}

\vspace{1cm}
\begin{titlepage}
\begin{flushleft}
       \hfill                      {\tt hep-th/0304009}\\
       \hfill                      HIP-2003-20/TH \\
       \hfill                      April, 2003\\
\end{flushleft}
\vspace*{3mm}
\baselineskip18pt
\begin{center}
{\Large {\bf On Free Field Realizations of Strings in BTZ}} \\
\vspace*{12mm}
{\large
Samuli Hemming\footnote{E-mail: Samuli.Hemming@helsinki.fi}} \\
\vspace{5mm}
{\em Helsinki Institute of Physics \\
P.O. Box 64\\
FIN-00014  University of Helsinki \\
Finland }
\vspace*{10mm}
\end{center}


\begin{abstract}

We discuss realizations of the $SL(2,R)$ current algebra in the 
hyperbolic basis using free scalar fields. It has been previously 
shown by Satoh how such a realization can be used to describe 
the principal continuous representations of $SL(2,R)$. We 
extend this work by introducing another realization that corresponds 
to the principal discrete representations of $SL(2,R)$. We show 
that in these realizations spectral flow can be interpreted 
as twisting of a free scalar field. Finally, we discuss how these 
realizations can be obtained from the BTZ Lagrangian. 

\end{abstract}

\end{titlepage}

\baselineskip16pt

\section{Introduction} 

$2+1$ dimensional anti-de Sitter space (AdS$_3$) is 
one of the recently studied examples 
of string theory in non-trivial backgrounds. 
Since AdS$_3$ is the group manifold of special linear 
transformations, it can be studied using a $SL(2,R)$ 
Wess--Zumino--Witten model. However, the physical spectrum of 
the $SL(2,R)$ WZW model seemed to contain states 
of negative norm. A resolution to this problem is to 
utilize an additional symmetry, known as spectral flow, in 
the theory. In the context of the $SL(2,R)$ WZW model, this symmetry 
was first discussed in \cite{Henningson,Hwang}, where it was noted 
that spectral flow was necessary for the modular invariance of the 
theory. In \cite{MO}, it was shown how the states with negative norm 
could be removed from the physical spectrum using spectral flow. 

One of the possible extensions of \cite{MO} is to study 
orbifolds of AdS$_3$ space-time. These include AdS$_3/Z_N$ orbifolds, 
which have been examined in \cite{Son,Martinec}. 
Another interesting case is the Ba\~{n}ados--Teitelboim--Zanelli 
(BTZ) black hole \cite{BTZ}, which can be obtained by quotienting 
AdS$_3$ by a boost. As opposed to pure AdS$_3$, specific features 
of the BTZ black hole include the appearance of space-time horizons 
and a region that is analogous to a space-like singularity. 
String theory on BTZ black holes based on the WZW model 
has been described in \cite{AK,Horowitz,Kaloper,NS}. 
In the context of spectral flow, the model has been 
analyzed by \cite{SHEKV,Troost}. Spectral flow in the fermionic 
sector has recently been discussed in \cite{Pakman}. 

In this letter, we consider a realization of the $SL(2,R)$ current 
algebra using free scalar fields. The realization is given in the 
hyperbolic basis which is the appropriate choice for a BTZ black hole. 
Similar investigation has been made 
by Satoh \cite{Satoh} (see also \cite{Bars}), but the 
discussion was limited to the 
principal continuous representations only. We will also clarify the 
role of the spectral flow in this setup. Spectral 
flow can be formulated as twisting of a free field in a basis where 
the generator of the Cartan subalgebra is diagonal. 
We will also discuss how the scalar field realization can be derived 
starting from the classical BTZ Lagrangian. We show that the values 
of the spectral flow parameters obtained in this way coincide with 
the results of a previous study \cite{SHEKV}. 

\section{$SL(2,R)$ Current Algebra} 

\subsection{Realizations in the Hyperbolic Basis} 
\label{sect_real} 

Most of this subsection is a review of \cite{Satoh}. 
The $SL(2,R)$ current algebra in the hyperbolic basis is 
expressed as 
\bea
J^2(z) \, J^\pm(z') &\sim& \frac {\pm i J^\pm(z')}{z-z'} \nonumber \\ 
J^+(z) \, J^-(z') &\sim& \frac {-k}{(z-z')^2} -\frac {2i J^2(z')}{z-z'} 
\label{alg1} \\ 
J^2(z) \, J^2(z') &\sim& \frac {k/2}{(z-z')^2} \nonumber 
\eea

We wish to construct a realization of (\ref{alg1}) using free fields. 
For the description of spectral flow, it is useful to work in a basis 
where the Cartan current $J^2$ is diagonal. A suitable realization is 
obtained when we 
introduce three bosonic scalar fields $X_a \ (a=0,1,2)$ obeying the 
following operator product expansion: 
\beq
X_a(z) \, X_b(z') \sim - \eta_{ab} \, {\rm ln} (z-z') 
\label{X_OPE}
\eeq 
The signature of the metric $\eta$ is chosen to be 
$\eta = {\rm diag} (-1,1,1)$. Then the currents satisfying (\ref{alg1}) 
can be constructed as in \cite{Satoh,IK}: 
\beq
\renewcommand{\arraystretch}{1.8}
\begin{array}{ccl} 
iJ^\pm &=& e^{\mp \sqrt{2/k} X_-} \, \partial \left( \sqrt{k/2} \, X_0 
\mp \sqrt{k'/2} \, X_2 \right) \\ 
iJ^2 &=& \sqrt{k/2} \ \partial X_1 
\end{array}
\label{cont_current} 
\eeq 
where the notations $X_\pm = X_0 \pm X_1$, $k'=k-2$ have been 
introduced. 

The energy-momentum tensor is obtained from the currents as follows: 
\beq
T = \frac 1{k'} \, \eta_{ab} \, (J^a J^b) = - \frac 12 \, 
\eta^{ab} \, \partial X_a \partial X_b + Q \,\partial^2 X_2  
\label{cont_tensor}
\eeq
The resulting central charge of the system is $c = 3k/k'$. 
We see that the energy-momentum tensor represents a system of three 
free fields, where the space-like field $X_2$ is coupled to the 
background charge $Q = \frac {-1}{\sqrt{2k'}}$. It turns out that 
this realization corresponds to the principal continuous 
representations of $SL(2,R)$. 

Next, one would like to know what are the vertex operators in this 
realization. One finds \cite{Satoh,IK} 
\beq
V_{j,J}(z) = {\rm exp} \left( i J \sqrt{2/k} \, X_- (z) + j \sqrt{2/k'} 
\, X_2(z) 
\label{cont_vertex}
\right) 
\eeq 
The relevant OPEs are then 
\beq 
J^2 (z) V_{j,J} (z') \sim \frac J{z-z'} \ V_{j,J} (z') \quad , \quad 
J^\pm (z) V_{j,J} (z') \sim \frac {J\mp ij}{z-z'} \ V_{j,J\pm i} (z') 
\label{vertex_alg}
\eeq 
The conformal weight of the vertex operator $V_{j,J}$ is related 
to the second Casimir of $SL(2,R)$ as 
\beq 
h_{V_{j,J}} = \frac {c_2}{k'} = \frac {-j(j+1)}{k-2} 
\label{dim} 
\eeq 

In the $SL(2,R)$ current algebra, there exist operators 
that contain only regular terms in their operator 
products with the currents. These so-called screening operators 
are given by \cite{Satoh}
\beq
\renewcommand{\arraystretch}{1.8}
\begin{array}{ccl} 
\eta^\pm &=& {\rm exp}\left( \pm \sqrt{k/2} \ X_0 - 
\sqrt{k'/2} \ X_2  \right) \\ 
S &=& \partial X_0 \ {\rm exp} \left( -\sqrt{2/k'} \ X_2 \right) 
\end{array} 
\eeq 
The expressions for the screening operators are determined up to 
constant factors and total derivatives. 

This construction of a realization of the $SL(2,R)$ current 
algebra using three scalar fields is not unique. 
We need another realization that corresponds to the principal 
discrete representations in order to fully describe the physical 
spectrum of the $SL(2,R)$ current algebra. In this letter, we point out 
that this realization is obtained by making the 
substitutions $X_2 \rightarrow -i X_0$, $X_0 \rightarrow i X_2$ in the 
previous formulae. Then the currents and the stress tensor become 
\bea
J^\pm &=& e^{\pm \sqrt{2/k} \ (X_1 - iX_2 )} \, \partial \left( \sqrt{k/2} 
\, X_2 \pm \sqrt{k'/2} \, X_0 \right) \nonumber  \\ 
iJ^2 &=& \sqrt{k/2} \ \partial X_1 \\ 
T &=& -\frac 12 \, \eta^{ab} \, \partial X_a \partial X_b + 
Q \, \partial^2 X_0 \nonumber 
\eea
with $Q = \frac i{\sqrt{2k'}}$. It should be noted that the field 
coupling to the background charge is now the time-like field $X_0$. 
The connection between this realization and the discrete 
representations will be established in the next section. 

The vertex operator corresponding to this realization is 
\beq
V_{j,J}(z) = {\rm exp} \left( -ij\sqrt{2/k'} \, X_0 (z) - iJ\sqrt{2/k} 
\, \left( X_1 (z) - i X_2 (z) \right) \right) 
\label{disc_vertex} 
\eeq 
and it has the same dimension and satisfies the same OPEs as in 
the previous realization (\ref{vertex_alg}), (\ref{dim}). 

\subsection{Representations of $SL(2,R)$} 

If a holomorphic field $X(z)$ is coupled to a background charge 
$Q$, the mode expansion for $X(z)$ can be written as 
\beq 
X (z) = \hat q - i \, ( \hat p - i Q ) \, {\rm ln} z 
+i \sum_{n \neq 0 } \frac { \alpha_n }{n} z^{-n} 
\eeq 
The corresponding Virasoro generators are then 
\beq
L_n = \frac 12 \, \sum_\ell \alpha_\ell \alpha_{n-\ell} 
+ i Q n \alpha_n + \frac 12 \, Q^2 \delta_{n,0}
\eeq 

The $SL(2,R)$ invariant vacuum state $\vac$ is invariant under 
the global conformal transformations, $L_{0,\pm 1} \vac = 0$. 
Using the realization (\ref{cont_tensor}), 
the vacuum state is found to be labeled by the background charge, 
\beq 
\vac = \left| \ p^0 = p^1 = 0, \ p^2 = iQ = -i/\sqrt{2k'} \ \right> 
\eeq 
States corresponding to the vertex operators (\ref{cont_vertex}), 
ie.~the primary states, are then 
\beq 
\left|\, j; J\right> = \lim_{z \to 0} V_{j,\, J} (z) \vac 
= \left| \ p^0 = p^1 = \sqrt{2/k} \ J, 
\ p^2 = i (Q-\sqrt{2/k'} \ j) \right> 
\eeq 
From this expression, we see that the relation between $j$ 
and the momentum component $p^2$ is 
\beq 
j = - \frac 12 + i \sqrt{k'/2} \ p^2 
\eeq 
Let us compare this with the $j$-values of the principal 
continuous representation $\calc$ of $SL(2,R)$. 
The unitary condition of $\calc$ is the following: 
\beq 
j= -\frac 12 + is, \ s \in {\bf R} 
\eeq
Hence we conclude that the realization (\ref{cont_vertex}) 
actually belongs to $\calc$, since the momentum component $p^2$ 
is real. 

However, we introduced another realization (\ref{disc_vertex}) 
where the background charge is coupled to a time-like field. 
This realization leads to a different set of vacuum and primary 
states: 
\bea 
\vac &=& \left| \ p^1 = p^2 = 0, \ p^0 = iQ = -1/\sqrt{2k'} \ \right> \\ 
\left|\, j; J\right> &=& \left| \ p^1 = -i p^2 = \sqrt{2/k} \ J, 
\ p^0 = i Q- \sqrt{2/k'} \ j \  \right> 
\eea 
The corresponding $j$-value is now 
\beq 
j = - \frac 12 - \sqrt{k'/2} \ p^0 
\eeq 
Now, we compare this with the $j$-values of the discrete representations 
$\cald^\pm$ of $SL(2,R)$. The unitarity condition 
\beq 
j < -\frac 12 
\eeq 
is satisfied if time is flowing forward, $p^0 > 0$. 

There exists a problem with the Hilbert space of the discrete 
representations. Namely, their spectrum is afflicted by ghosts 
(states with negative norm). The ghosts appear in the discrete 
representations when $j<-k/2$. The spectrum could be truncated by 
hand, but doing so would create a lower bound on the second Casimir 
$c_2 = -j(j+1)$, which is related to the mass of a state. 
In string theory, such a bound would be considered artificial. 
A way out of this conundrum was given in \cite{MO} using the 
spectral flow symmetry of the theory. 

The fact that $J^\pm$ shifts $J$ 
by $\mp ij$ as in equation (\ref{vertex_alg}) might first appear 
puzzling, but this is actually a feature of the hyperbolic basis 
\cite{KMS,NS,SHEKV}. The operator $J^2$ now corresponds to a 
non-compact direction of the target space, and has a continuous 
spectrum. Therefore the states in the Hilbert space must be defined 
as 
\beq
\left| \Phi \right> = \int_{-\infty}^{\infty} dJ \ \Phi(J) 
\left|\, j;J\right> 
\eeq
The action of the currents $J^2,J^\pm$ on the states are 
\bea 
J^2_0 \left| \Phi \right> &=& \int_{-\infty}^{\infty} dJ \ 
J \, \Phi(J) \left|\, j;J\right> \nonumber \\ 
J^+_0 \left| \Phi \right> &=& \int_{-\infty}^{\infty} dJ \ 
f(J) \, \Phi(J-i) \left|\, j;J\right> \\ 
J^-_0 \left| \Phi \right> &=& \int_{-\infty}^{\infty} dJ \ 
g(J+i) \, \Phi(J+i) \left|\, j;J\right> \nonumber 
\eea 
The functions $f(J)$, $g(J)$ correspond to matrix elements of 
$J^+_0$, $J^-_0$. 

\subsection{Twisting and Spectral Flow} 

We now turn to discuss spectral flow in the free field realization. 
Our goal is to show how spectral flow can be interpreted as twisting 
of the field $X_1$ in the realizations introduced in section 
\ref{sect_real}. The BTZ space-time has the topology of a solid 
cylinder with one compact space-like direction. Twisted sectors of 
the theory are generated by winding the string around the compact 
dimension. It will be shown in section \ref{Lagr_appr} that the 
winding of the string can be accomplished by twisting the field $X_1$. 

In the presence 
of rotation, the mode expansion for a scalar field compactified on 
a circle of radius ${\cal R}= {\rm Re} \ R=\frac 12(R+\bar R)$ 
becomes 
\beq 
X_1 (z,\zbar) = \hat q^1 - \frac i2 \, ( \hat p^1 + m R) \, {\rm ln} z 
- \frac i2 \, ( \hat p^1 - m \bar R) \, {\rm ln} \zbar 
+i \sum_{n \neq 0 } \frac 1n \left( \alpha_n^1 z^{-n} 
+ \bar \alpha_n^1 \zbar^{-n} \right) 
\label{twist_exp}
\eeq 
The imaginary part of $R$ is related to the intrinsic angular momentum 
of the rotating target space. If the system is non-rotating, 
$\bar R = R$. The integer $m$ measures the winding of the field 
$X_1$ around the compact dimension. Also, the momentum becomes quantized, 
$p^1 = n/{\cal R}$. 

The mode expansion (\ref{twist_exp}) tells that twisting acts as 
shifting of the momentum $p^1$. Effectively, twisting is 
\beq 
\alpha^1_0 \rightarrow \alpha^1_0 -  \sqrt{2k} \, w \ , \quad 
\bar \alpha^1_0 \rightarrow \bar \alpha^1_0 + \sqrt{2k} \, \bar w 
\label{spf} 
\eeq 
where we have introduced $w = - mR/\sqrt{2k}$, 
$\bar w = -m \bar R / \sqrt{2k}$ for convenience. 
Consequently, the currents transform under (\ref{spf}) as 
\beq
\begin{array}{ccccccccccc}
J^\pm (z) &\rightarrow& \tilde J^\pm (z) &=& J^\pm (z) \, z^{\pm iw} 
&, & 
\bar J^\pm (\zbar) &\rightarrow& \tilde {\bar J}^\pm (\zbar) &=& 
\bar J^\pm (\zbar) \, \zbar^{\pm i\bar w} \\ 
J^2 (z) &\rightarrow& \tilde J^2 (z) &=& J^2 (z) + \frac {kw}{2z} &, & 
\bar J^2 (\zbar) &\rightarrow& \tilde {\bar J}^2 (\zbar) &=& 
\bar J^2 (\zbar) - \frac {k\bar w}{2\zbar} 
\end{array} 
\eeq
and the transformation of the energy-momentum tensor is the following: 
\beq
\renewcommand{\arraystretch}{1.8} 
\begin{array}{ccc} 
T(z) &\rightarrow& \tilde T(z) = T(z) + \frac wz \, J^2(z) + 
\frac {k w^2}{4z^2} \\ 
\bar T(\zbar) &\rightarrow& \tilde {\bar T}(\zbar) = \bar T(\zbar) 
- \frac {\bar w}{\zbar} \, \bar J^2(\zbar) + 
\frac {k \bar w^2}{4\zbar^2} 
\end{array} 
\eeq 
Notice that the this transformation is independent of the 
representations $\calc$ and $\cald^\pm$.\footnote{
Note that also the screening charges remain invariant.}

The transformed currents $\tilde J^a$ satisfy the $SL(2,R)$ algebra 
(\ref{alg1}) and the Virasoro algebra. Thus the transformation 
(\ref{spf})  generates a new representation of the current 
algebra, which should be taken into account when considering the 
Hilbert space of the theory. 

In terms of modes, the transformation reads 
\beq
\renewcommand{\arraystretch}{1.5}
\begin{array}{cclcccl} 
J^\pm_n &\rightarrow& J^\pm_{n\pm iw} & & 
\bar J^\pm_n &\rightarrow& \bar J^\pm_{n\pm i\bar w} \\ 
J^2_n &\rightarrow& J^2_n + \frac k2 w \, \delta_{n,0} & & 
\bar J^2_n &\rightarrow& \bar J^2_n - \frac k2 \bar w \, \delta_{n,0} \\ 
L_n &\rightarrow& L_n + w J^2_n + \frac k4 w^2 \delta_{n,0} & & 
\bar L_n &\rightarrow& \bar L_n - \bar w \bar J^2_n + 
\frac k4 \bar w^2 \delta_{n,0} 
\end{array} 
\eeq 
This result is in exact agreement with \cite{SHEKV}. Hence we identify 
the transformation (\ref{spf}) as spectral flow in the free field 
realization. Because the transformation (\ref{spf}) is nothing but 
twisting, we conclude that spectral flow in the free field 
realization is twisting of the field $X_1$. A remaining task is to find 
out what is the compactification radius $R$. 
We will discuss this in section \ref{Lagr_appr}. 

The importance of spectral flow comes from the fact that it can be 
used to eliminate the appearance of ghosts in the discrete representations 
of $SL(2,R)$. In general, the spectral flowed discrete representations 
are related by 
\beq
\cald^{\pm,w}_j = \cald^{\mp,\tilde w = 1\pm w}_{-k/2 -j} 
\eeq
It was shown by Maldacena and Ooguri \cite{MO} that spectral flow 
naturally imposes the limit $-1/2 > j > -(k-1)/2$ on the values 
of $j$. Within this range of $j$, ghosts do not appear in the 
spectrum of the discrete representations. If spectral flow is taken 
to be a symmetry of the theory, the theory will be free of ghosts. 

\section{Lagrangian Approach} 
\label{Lagr_appr}

\subsection{BTZ Coordinates} 

The aim here is to show explicitly how the free fields $X_a$ 
introduced in the previous section are related with the fields appearing 
in the sigma model form of the BTZ Lagrangian. For a correct description 
of the $SL(2,R)$ current algebra at the quantum level, one needs to 
resolve some subtleties arising from the transformation of the 
classical Lagrangian. For the AdS$_3$ case, similar discussion has 
been given in \cite{GKS,HOS,GN}. 

The BTZ metric \cite{BTZ} for a $2+1$ dimensional black hole is 
\beq
ds^2 = - \ \frac {(r^2-r_+^2)(r^2-r_-^2)}{r^2} dt^2 
+ \frac {r^2 \ dr^2}{(r^2-r_+^2)(r^2-r_-^2)} 
+r^2 \, \left( d\theta - \frac {r_+ r_-}{r^2} dt^2 \right)^2 
\label{btz_metric}
\eeq 
Here $r_-$ and $r_+$ are the inner and outer horizons, respectively. 
The angular coordinate has periodicity $\theta \sim \theta + 2\pi$. 
The time coordinate is usually taken to be non-compact, 
$-\infty < t < \infty$. 

The mass and the angular momentum of the black hole are given in 
terms of the inner and outer horizon, 
$M_{BH}= r_+^2 + r_-^2$ and $J_{BH}= 2r_+ r_-$. 
For a non-rotating black hole, the inner horizon goes to zero: 
$r_- = 0$. 

To proceed with the analysis, we transform the metric 
(\ref{btz_metric}) into a more convenient form. We make an 
analytic continuation to Euclidean time $t_E = it$ and make 
a coordinate transformation 
\bea 
\gamma &=& \sqrt{\frac {r^2-r_+^2}{r^2-r_-^2}} \ e^{ (r_+ -r_- ) 
(\theta - it_E ) } \nonumber \\ 
\bar \gamma &=& \sqrt{\frac {r^2-r_+^2}{r^2-r_-^2}} \ e^{ (r_+ +r_- ) 
(\theta + it_E ) } \\ 
\phi &=& -\frac 12 \ {\rm ln} \left( \frac {r_+^2-r_-^2}{r^2-r_-^2} \right) 
- (r_+ \theta + i r_- t_E ) \nonumber 
\eea 
The resulting metric is the Poincare patch for Euclidean AdS$_3$. 
The Lagrangian describing string propagation on this manifold 
can be written in the sigma model form as 
\beq
L = k ( \partial \phi \bar \partial \phi + e^{2 \phi} 
\partial \bar \gamma \bar \partial \gamma ) 
\label{ads_euc} 
\eeq 
The level number $k$ has been included for the correct normalization 
of the path integral. The periodicity of the BTZ angular coordinate 
$\theta$ translates into the following periodic identifications for 
$\gamma,\bar \gamma,\phi$: 
\beq 
\gamma (z) \sim \gamma (z) e^{2\pi \Delta_-} \ , \quad 
\bar \gamma (\zbar) \sim \bar \gamma (\zbar) e^{2\pi \Delta_+} \ , \quad 
\phi (z,\zbar) \sim \phi (z,\zbar) - 2\pi r_+ 
\label{peri}
\eeq 
where $\Delta_\pm$ is given in terms of the 
inner and outer horizon as $\Delta_\pm = r_+ \pm r_-$. 

A common trick is to rewrite the Lagrangian by introducing new 
fields $\beta, \bar \beta$. The original Lagrangian (\ref{ads_euc}) 
can then be recovered from 
\beq
L = k \partial \phi \bar \partial \phi - \beta \bar \partial \gamma 
- \bar \beta \partial \bar \gamma - \frac 1k \beta \bar \beta 
e^{-2 \phi} 
\eeq 
by integrating over the fields $\beta, \bar \beta$. At quantum level, 
however, the above transformation leads to an anomalous factor 
in the functional measure. After evaluation of this factor 
(for details, see \cite{HOS}), one gains the effective Lagrangian: 
\beq 
L = k' \partial \phi \bar \partial \phi - \frac {\phi}4 
\sqrt{h} R^{(2)} - \beta \bar \partial \gamma 
- \bar \beta \partial \bar \gamma - \frac 1k \beta \bar \beta 
e^{-2 \phi } 
\label{L_bg}
\eeq 
Here $R^{(2)}$ is the scalar curvature in two dimensions. It is 
useful to think of the interaction term 
$L_{int}= - \frac 1k \beta \bar \beta e^{-2 \phi}$ as a screening 
current. In the limit $\phi \rightarrow \infty$, it can be treated 
perturbatively.  

\subsection{Currents} 

The free part of the Lagrangian (\ref{L_bg}) leads to the Wakimoto 
free field realization \cite{Wakimoto} of the $SL(2,R)$ current 
algebra. In this realization, 
coordinates $\beta, \gamma$ ($\bar \beta, \bar \gamma$) constitute 
a holomorphic (antiholomorphic) system of bosonic ghosts. Their OPE 
is given by 
\beq
\beta (z) \gamma (z') \sim \frac 1{z-z'} 
\eeq 
and a similar relation holds for the antiholomorphic fields. 
In the hyperbolic basis, the holomorphic currents of the $SL(2,R)$ 
current algebra are: 
\bea
i J^+(z) &=& \beta (z) \nonumber \\ 
i J^-(z) &=& ((\gamma \gamma)\beta) (z) +\sqrt{2k'}\ 
(\gamma \partial \phi)(z)  + k\, \partial \gamma (z) \\ 
i J^2(z) &=& (\gamma \beta) (z) + \sqrt{k'/2}\ \partial \phi(z) \nonumber 
\eea 

Now, we introduce three scalar fields $X_a$ and (re-)bosonize the 
($\beta ,\gamma$) system. The OPEs for the fields $X_a$ have 
been defined in (\ref{X_OPE}). The transformation between 
$\beta ,\gamma ,\phi$ and $X_a$ is 
\bea
\beta (z) &=& e^{-\sqrt{2/k}\, X_-(z)} \partial \left( \sqrt{k/2} \ 
X_0 (z) - \sqrt{k'/2} \ X_2 (z) \right) \nonumber \\ 
\gamma (z) &=& e^{\sqrt{2/k}\, X_-(z)} \\ 
\phi (z) &=& 1/\sqrt{2k'} \ X_2 (z) - 1/\sqrt{2k} \ X_- (z) \nonumber 
\eea 
Not surprisingly, the resulting currents and the energy-momentum 
tensor coincide with the ones given in the previous section 
(\ref{cont_current}), (\ref{cont_tensor}). 

Next we consider the periodicity of the coordinates. From 
(\ref{peri}) we find that 
\beq  
X_- (z) \sim X_- (z) + \pi \sqrt{2k} \ \Delta_- \quad , \quad 
\tilde X_- (\zbar ) \sim \tilde X_- (\zbar ) + \pi \sqrt{2k} \ \Delta_+
\eeq 
In the spirit of the previous section, we require that 
the twisted sectors of the system are generated by twisting of 
the field $X_1$. The respective periodicities for holomorphic and 
antiholomorphic part of the coordinate $X_1$ are then 
\beq  
X_1 (z) \sim X_1 (z) - \pi \sqrt{2k} \ \Delta_- \quad , \quad 
\tilde X_1 (\zbar ) \sim \tilde X_1 (\zbar ) - \pi \sqrt{2k} \ \Delta_+
\eeq 
In this setup, $X_0$ and $X_2$ are not periodic.\footnote{
If we considered a single cover of the $SL(2,R)$ group 
manifold, the time-like coordinate $X_0$ would be periodic.} 
The same periodicities are obtained for the discrete representations. 

Comparing the mode expansion (\ref{twist_exp}) and the spectral flow 
transformation (\ref{spf}), we find 
\beq
w=m\Delta_- \quad , \quad \bar w = m \Delta_+
\eeq
A similar relation was found in \cite{SHEKV} using classical 
considerations. 

\section{Summary} 

We have examined a realization of the $SL(2,R)$ current algebra in 
the hyperbolic basis using free scalar fields. The energy-momentum 
tensor of the model has a simple form, where one of the scalar fields 
couples to a gravitational background charge. 
We have shown that the realizations belong to the principal continuous 
representations $\calc$ if the field coupling to the background 
charge is space-like, and to the principal discrete representations 
$\cald^\pm$ if the coupled field is time-like. This has a nice 
analogy in \cite{MO}, where it was found that 
long string states generated by spectral flow of space-like geodesics 
correspond to principal continuous representations, 
and short string states generated by spectral flow of 
time-like geodesics correspond to principal discrete representations. 

We demonstrated how spectral flow could be interpreted as 
twisting of a free scalar field in this model. Using the BTZ Lagrangian 
as a starting point, we reviewed the construction of the free field 
realization of the $SL(2,R)$ current algebra. The periodic conditions 
of the BTZ coordinates imply that only a discrete set of values is 
allowed for the spectral flow parameters. These values also agree 
with \cite{SHEKV}. 

\bigskip 

\noindent 
{\bf \large Acknowledgments} 

\bigskip 

\noindent 
The author would like to thank E.~Keski-Vakkuri and S.~Kawai 
for useful discussions.

\end{document}